\documentstyle[epsfig]{mn}
\begin{document}

\title
[GRB afterglows and cosmic star formation]
{Extinction of gamma-ray burst afterglows as a diagnostic of the
location of cosmic star formation}
\author[E. Ramirez-Ruiz, N. Trentham \& A.\,W. Blain] 
{Enrico Ramirez-Ruiz, Neil Trentham and A.\,W. Blain\\ 
Institute of Astronomy, Madingley Road, Cambridge, CB3 0HA.}
\maketitle 

\begin{abstract} 
{ 
The properties of gamma-ray bursts (GRBs) and their
afterglows are used to investigate the location of star formation
activity through the history of the Universe.
This approach is motivated by the following:
(i) GRBs are thought to be associated
with the deaths of massive stars and so the GRB
rate ought to follow the formation rate of massive stars, 
(ii) GRBs are the last phase of the evolution of these
stars, which do not live long enough to
travel far from their place of birth,
and so GRBs are located where the stars formed,
and (iii) GRB afterglows occur over both X-ray
and optical wavelengths, and so the differential effects of dust
extinction between 
the two wavebands can reveal whether or not
large amounts of dust are present in galaxies
hosting GRBs. 
Recent  evidence suggests that a significant fraction of stars in the Universe
formed in 
galaxies that are bright at rest-frame submillimetre (submm) and
infrared wavelengths rather
than at ultraviolet wavelengths; we estimate about three quarters
of the star formation in the Universe occurred in the submm-bright mode. 
High-redshift submm-selected galaxies are thought to have 
properties similar to local ultraluminous
infrared galaxies (ULIGs) like Arp 220, based on the
concordance between their luminosities and spectral energy
distributions.
If this is the case, then GRBs in submm-bright
galaxies should have their optical afterglows extinguished by internal
dust absorption, but only very few should have their 2$-$10 keV
X-ray afterglows obscured.
The value of three quarters that we quote is
marginally consistent with observations of GRBs: $60 \pm 15$ per cent of 
GRBs have no detected optical afterglow, whereas  almost all have an X-ray
afterglow. A more definitive statement could be made with observations of
soft X-ray afterglows (0.5$-$2 keV), in which extinction should be
severe for  GRBs located in
submm-bright galaxies with gas densities similar to those
in local ULIGs.  If the X-ray afterglows disappear at soft X-ray
wavelengths in a large number of GRBs, then this would provide strong
evidence that much of the star-formation in the Universe is heavily
obscured. Far-infrared and submm follow up studies of the hosts of GRB
would reveal this population. We expect about 20\,per cent of GRB hosts 
to be detectable using the SCUBA camera at the JCMT after several hours
of integration. 
}
\end{abstract} 


\section{Introduction}

\begin{figure*}
\begin{center}
\vskip-1mm
\psfig{file=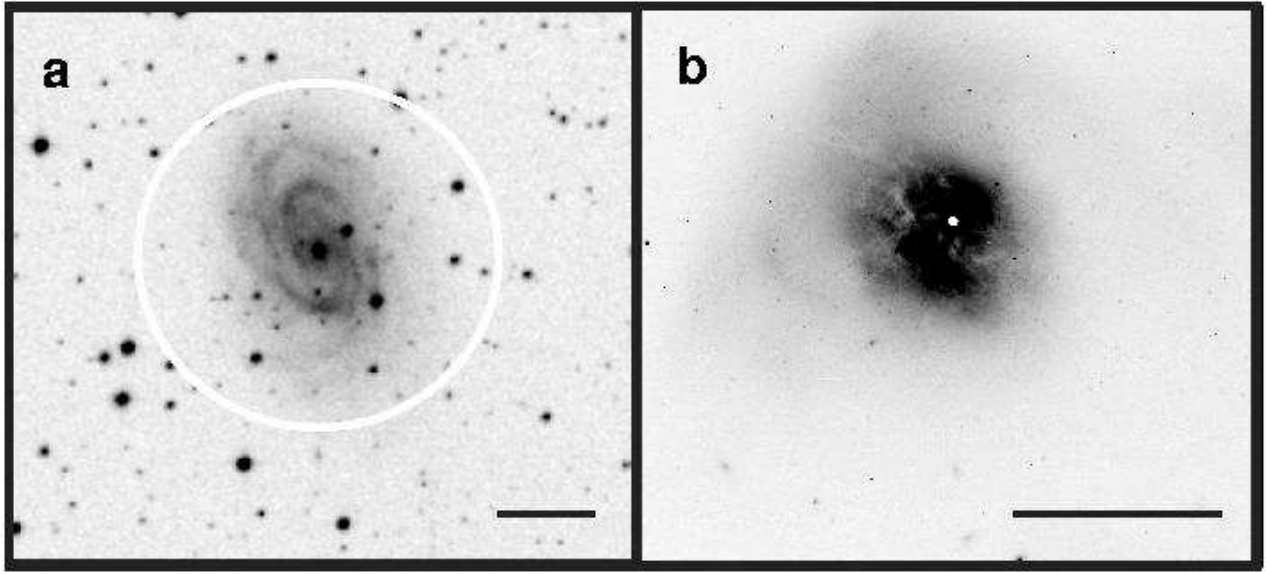, width=0.95\textwidth}
\end{center}
\caption{
Two extreme examples of local galaxies, representing star formation 
happening in the UV and submm modes and the consequent effects on 
extinction of the radiation produced by the young stars. The bars represents 
10\,kpc in the rest frame of the galaxy.
(a) NGC 4603 (image from the Palomar Observatory Sky Survey)
is a luminous Sc galaxy that is bright at UV and blue wavelengths.
The circle represents the 90\,per cent isophote, indicating that the
star-formation is occurring over the whole galaxy.
(b) Arp 220 (image from {\it HST} WFPC2) is a ULIG that is very bright at 
far-infrared wavelengths, but very faint at ultraviolet wavelengths 
(Meurer et al.~2001). The white dot represents the size of the 
far-infrared-emitting blackbody nucleus (Solomon et al.~1997), indicating 
that the star-formation activity is concentrated near the centre of
the galaxy. The far-infrared-emitting blackbody nucleus represents 
90\,per cent of the total bolometric emission and hence the white dot
is the equivalent of the  90\,per cent isophote shown in  panel (a). 
}
\end{figure*}
 
One of the major recent achievements in extra-galactic astronomy has
been an estimate of the history of the cosmic star
formation rate, using two types of complementary measurements. First, 
optical and near-infrared observations can be made of distant galaxies.
The rest-frame ultraviolet (UV) or optical luminosities of these galaxies
can then be converted into star formation rates and summed over a 
representative sample of galaxies in each redshift interval to derive the 
form of the comoving star formation rate of the Universe
as a function of redshift, as is conventionally presented in the
Madau plot (Lilly et al.~1996, Madau et al.~1996;
see Somerville, Primack \& Faber 2000 or Blain 2001 for recent compilations). 
Secondly, submm (Blain et al.~2000) 
and far-infrared (Kawara et al.\ 1998; Puget et al.\ 1999)
measurements of source
counts  and the extra-galactic background light 
(Fixsen et al.\ 1998; Hauser et al.~1998;
Schlegel et al.\ 1998) can be used to
infer the amount of dust-enshrouded high-redshift star-formation
activity. The redshift distribution of sources
contributing to these source counts and backgrounds is poorly known;
nevertheless the constraints are tight enough that reasonable progress
has been made in constructing a far-infrared/submm Madau plot 
(Blain et al.\ 1999b, Blain 2001).

At redshifts $z>3$ the most luminous UV-bright
galaxies are the Lyman break galaxies, identified by the Lyman break
technique of Steidel et al. (1996) and
the most luminous submm
galaxies are the galaxies seen by the Submillimetre Common User
Bolometer Array (SCUBA) on the James Clerk Maxwell Telescope
(Smail, Ivison \& Blain 1997, Barger et al.~1998, Hughes et al.~1998,
Eales et al.~1999). Both the Lyman break (Steidel et al.~1999) 
and some SCUBA (Ivison et al.~1998, 2000) galaxies are now being studied 
in some detail.

The most striking conclusion from these studies are that the Lyman break 
and SCUBA galaxies appear to be distinct populations of objects. 
The Lyman break galaxies are faint at submm wavelengths (Ouchi et al.~1999; 
Peacock et al.~2000; Sawicki 2000), although they certainly undergo 
some internal dust extinction (Calzetti et al.~1997; Steidel et al.~1999). The 
SCUBA galaxies tend to be faint at rest-frame ultraviolet wavelengths
(Ivison et al.~2000; Smail et al.\ 2000, 2001) given their huge bolometric
luminosities. Both sets of sources contain extremely luminous galaxies. 
In fact the 850-$\mu$m background is composed at the 50\% level by 
sources with bolometric luminosities above 10$^{12}$\,L$_{\odot}$
(Blain et al.\ 1999a)
\footnote{Luminosities corresponding to 850-$\mu$m flux densities of
1\,mJy or greater.}, comparable to those of
local ultraluminous infrared galaxies  (ULIGs; Sanders \& Mirabel
1996). These conclusions are perhaps  
suggestive of an analogy with the nearby Universe, in which the most 
powerful star-forming galaxies are either UV-bright late-type giant Sc
galaxies like NGC 4603 (see Fig.\,1a) or infrared-bright ULIGs 
like Arp 220 (see Fig.\,1b), which are extremely faint at rest-frame UV 
wavelengths (Meurer et al.~2001), presumably due to internal extinction.

It is therefore plausible  that the star formation in the Universe happened
predominantly in two distinct modes -- UV and submm.  If the analogy
with the nearby Universe is valid (for the SCUBA galaxies this is
still speculative since very-high-resolution submm maps do not yet exist), 
then we expect the star formation to occur in either relatively diffuse and 
unobscured UV-bright star-forming galaxies or in very compact and heavily 
obscured (the internal extinction to the central power source in Arp 220 is 
greater than 30\,mag; Genzel et
al.~1998, Scoville et al.~1998, Shioya, Trentham \& Taniguchi 2001)
submm-bright ULIGs. 

One important remaining question is: how much of the star formation
that has happened over the age of the Universe occurred in each mode?
It appears that most of the energy is generated 
in the submm mode (Blain et al.\ 1999b,c).  However, this does not 
necessarily mean that most of the star formation in the Universe is 
happening in this mode, since (i) the stellar initial mass function (IMF)
may be high-mass biased in the SCUBA galaxies (Blain et al.~1999b),
resulting in more energy generation per unit mass of stars formed; (ii) 
there may be a substantial contribution to the energy output of SCUBA
galaxies from active galactic nuclei (AGNs). Given the huge
line-of-sight extinctions to the galaxy centres, it is not
possible to determine the power sources unambiguously; (iii)
there may be considerable star formation activity in UV-bright galaxies
that is hidden by dust, but in which the extinction is not sufficiently
great to generate SCUBA galaxies (Adelberger \& Steidel 2000). What we 
see directly in the UV would then only be a fraction of the total star 
formation activity in those galaxies. Both the total amount of obscured 
star formation in the Lyman break galaxies and the conversion factor from 
UV luminosity to a star formation rate are uncertain (Pettini et al. 2001).  

We now attempt to address this question by studying the extinction 
towards gamma-ray bursts (GRBs) and their afterglows. Several 
arguments suggest such an approach. First, GRBs are thought to be associated
with the deaths of massive stars and so their formation rate ought to 
trace that of the massive stars responsible for the observed UV and 
far-infrared luminosities of both the Lyman break and SCUBA galaxies 
(Totani 1997, Blain \& Natarajan 2000; Lamb \& Reichart 2000; Ramirez-Ruiz,
Fenimore \& Trentham 2001). Secondly, GRBs are the final phase of the 
evolution of these massive stars. These stars do not live long enough to 
travel far from their place of birth, and so GRBs should be located where 
the stars formed. Thirdly, GRB afterglows are seen in both the optical
and X-ray wavebands, and so we can hope to exploit the differential effects 
of extinction between the two wavebands to test whether large amounts of 
dust and gas are present in the galaxies hosting the GRBs. We can then 
infer how many of the GRBs occurred in galaxies with both
moderate and considerable amounts of dust, likely to be Lyman break and 
SCUBA galaxies respectively.

To elucidate the total amount of obscured star formation that happened
over the age of the Universe, we investigate the
direct observational constraints on the UV/submm fraction
of cosmic star formation  in more detail in Section 2.  We outline
properties of GRBs and their afterglows that
make them suitable for this kind of analysis in Section 3, and
in Section 4 we describe how dust might affect the
emergent X-ray and optical radiation from GRBs and their
afterglows. 
The constraints on the UV/submm
fraction from analysis of the extinction of GRB afterglows
are presented in Section 5. 
We present some caveats in 
Section 6, and
in Section 7 we summarise our results and outline future prospects.  
We assume $H_0 = 65\,\, {\rm km} {\rm s}^{-1}
{\rm Mpc}^{-1}$,
$\Omega_{\rm m}=0.3$,
$\Omega_{\Lambda}=0.7$.

\section{The location of cosmic star formation } 

\begin{figure}
\begin{center}
\vskip-4mm
\psfig{file=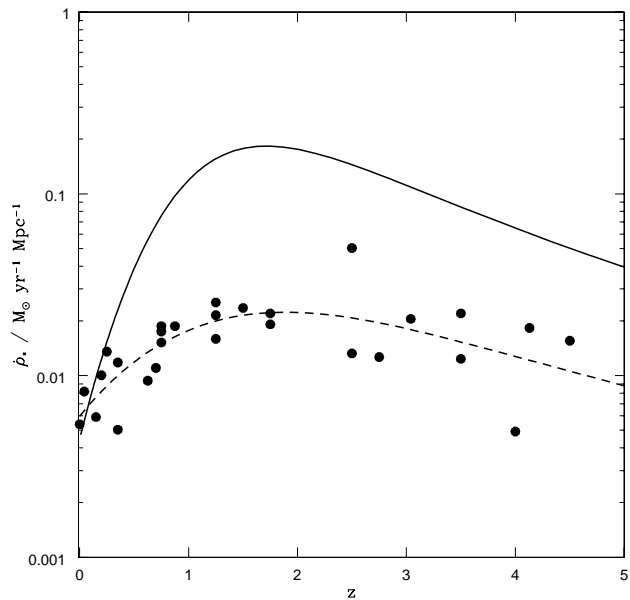, width=8.65cm}
\end{center}
\vskip-3mm
\caption{
The comoving star-formation rate of the Universe as a function of redshift.  
The points represent the optical/UV data compiled by Somerville, Primack 
\& Faber (2000) with no corrections for dust extinction; the dashed line 
represents a 4th-order polynomial fit to this data -- the UV--optical Madau
plot. The solid line represents the total (UV and submm) Madau plot, from the
models of Blain et al.\ (1999a,b) and Blain (2001), assuming no AGN 
contribution to the far-infrared and submm backgrounds and counts. The 
stellar IMF of Kroupa, Tout \& Gilmore (1993) is assumed.  
}
\end{figure}

A compilation of measurements of the UV star formation rate history
is presented in Figure 2 (Somerville, Primack \& Faber 2000), along
with a model curve that shows 
the total (UV and  submm) star formation rate history that has been derived
from a fit to the far-infrared and submm counts and
backgrounds (Blain 2001).  Since the curve is so much higher than the points
it is immediately apparent that most star formation has 
happened in the submm mode, subject to the caveats discussed in the last
section.  Given that a significant fraction of the submm-wave background is 
originated in luminous SCUBA galaxies (Blain et
al.~1999a), this suggests that most of the high-mass star-formation
activity in the  Universe is heavily obscured. 

To investigate this in more detail we now compute $F_{\rm submm}$, the 
fraction of the total star formation activity over the age of the Universe 
that has taken place in the submm mode, and compare with the fraction of 
GRBs for which afterglows might have been expected to be extinguished 
by an amount greater than the average line-of-sight optical depth to a 
star forming in a SCUBA galaxy.
We define
\begin{equation} 
F_{\rm submm} ={ 
{ {\int_{\infty}^{0} \dot{\rho}_{\rm submm} (z) 
  {{ {\rm d}t}\over{ {\rm d}z}} { {\rm d}z} } 
}\over
{ \int_{\infty}^{0} \dot{\rho}_{\rm TOT} (z) 
  {{ {\rm d}t}\over{ {\rm d}z}} { {\rm d}z}   
}}, 
\end{equation}
where
\begin{equation} 
\dot{\rho}_{\rm TOT} (z) = \dot{\rho}_{\rm submm} (z) 
+ \dot{\rho}_{\rm UV} (z) 
\end{equation}
is shown by the solid line in Fig.\,2, while $\dot{\rho}_{\rm UV}$ is
shown by the  dashed line. Here $t(z)$ is the age-redshift relation,
determined by the adopted cosmology. 
In order to evaluate this integral, we need to derive
the instantaneous submm fraction 
\begin{equation} 
f_{\rm submm}(z) = {{\dot{\rho}_{\rm submm} (z)
}\over{ \dot{\rho}_{\rm TOT} (z)} }
\end{equation}
as a function of redshift. 

\begin{figure}
\begin{center}
\vskip-4mm
\psfig{file=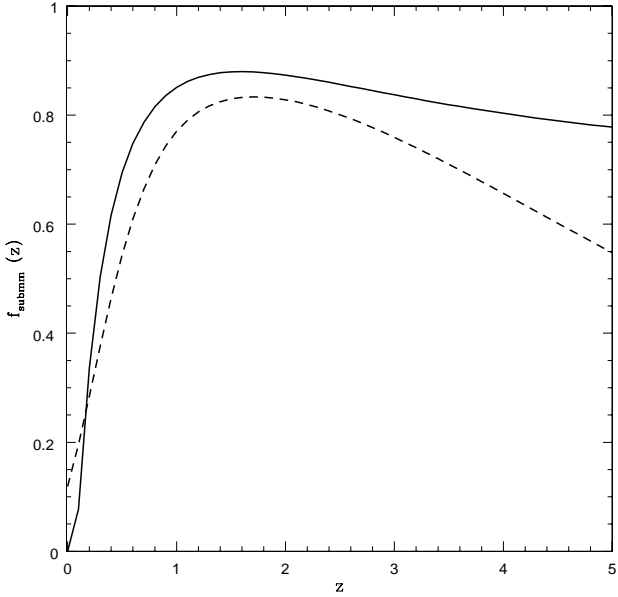, width=8.65cm}
\end{center}
\vskip-3mm
\caption{
The instantaneous submm fraction as a function of redshift
$f_{\rm submm} (z)$ computed as described in the text
by method A (solid line) or method B (dashed line)}
\end{figure}

We can compute $f_{\rm submm}(z)$ in two different ways, both shown in
Figure 3.  First, in Method A, we simply evaluate the ratio 
$\dot{\rho}_{\rm submm}(z) / \dot{\rho}_{\rm TOT} (z)$ from Figure 2. 
Secondly, in Method B, we assume that {\bf all} the star-formation activity
in objects above some threshold far-infrared luminosity $L_{\rm FIR,c}$ is 
obscured and occurs in the submm mode, whereas at luminosities
less than the threshold, all the star-formation activity is unobscured and 
visible in the UV mode. This assumption is motivated by the observation 
(Rieke \& Lebofsky 1986) that for low-redshift galaxies, the blue luminosity 
is proportional to the far-infrared luminosity for luminosities below some 
critical value (6.31 $\times 10^{10}$ L$_{\odot}$ in terms of the 60-$\mu$m 
luminosity L$_{60}$). If we assume that this 
value does not depend on  
redshift, then
\begin{equation}
f_{\rm submm}(z) =   
{ {\int_{6.31 \times 10^{10} {\rm L}_{\odot}}^{\infty}
L_{60} \phi(L_{60},z)\; {\rm d}L_{60}
}\over
{\int_{0}^{\infty}
L_{60} \phi(L_{60},z)\; {\rm d}L_{60}
}}
\end{equation}
where (Blain 2001)
\begin{equation}
\phi(L_{60},z) = 
\phi(L_{60}/g(z),0) 
,\end{equation}
\begin{equation}
g(z) = 10.42 \, (1+z)^{1.5} 
{\rm sech}^{2}[2.2 \ln (1+z) - 1.84],  
\end{equation}
\begin{equation} 
\phi_z (L_{\rm 60}) ={1\over{L_{\rm 60} \, \rm{ln} 10}} \left( {{L_{\rm 60}} 
\over {L_{\rm 60}^*}} \right)^{1 - \alpha} e^{h(L_{\rm 60})},
\end{equation}
\begin{equation} 
h(L_{\rm 60})= - {1\over{2 \sigma^2}}
\log_{10}^2 \left( 1 + {{L_{\rm 60}} \over {L_{\rm 60}^* }}\right),
\end{equation}
and ($\alpha$, $\sigma$, $L_{\rm 60}^*$) = (1.1, 0.724, 7.0 $\times$ 10$^8$
L$_{\odot}$), as for the local Universe (Saunders et al.~1990, adapted
to our cosmology).\\

The  computed values of $f_{\rm submm}(z)$ are presented in
Figure 3.  The methods agree reasonably well, except at high
redshift where the data is rather sparse and uncertain, and so the
models of $\dot{\rho}_{\rm TOT} (z)$ are poorly 
constrained. 
 
The global submm fractions $F_{\rm submm}$ are then
0.82 $\pm$ 0.10 (method A) or 0.74 $\pm$ 0.12 (method B). 
This quantifies a statement we made earlier: {\bf about three quarters
of the star formation in the Universe happened in the submm mode}.

As the interpretation of the rest of this paper depends
on the accuracy of this statement, it is worth examining
the caveats we listed in the previous section in some detail.

\begin{enumerate} 
\item IMF dependence.  If the IMF is high-mass biased in the SCUBA
galaxies, we need to multiply $F_{\rm submm}$ by
\begin{equation}
\zeta = {\xi \over 1 + F_{\rm submm}(\xi - 1)}
\end{equation}
to get the total fraction by mass of the stars in the
local Universe that formed in the submm mode, where
\begin{equation} 
\xi = { \int \Phi_{\rm S} {\rm m dm} \over  \int \Phi_{\rm UV} {\rm m dm}}\;\;,
\end{equation}
$\Phi_{\rm S}$ is the IMF of stars forming in a SCUBA galaxy
and $\Phi_{\rm UV}$ is the IMF of stars forming in a UV-bright galaxy.
If $\Phi_{\rm S}$ is truncated at lower masses $\zeta$  can be
significantly less than  1.
In this work however, we are concerned with $F_{\rm submm}$ and not
$\zeta F_{\rm submm}$ since GRBs are expected to be generated only by
the highest mass stars. 
\item AGN dependence.  If the SCUBA sources are powered predominantly
by AGN, then our derived value of $F_{\rm submm}$ should be multiplied
by a factor $(1-f_{\rm AGN})/(1-f_{\rm AGN}F_{\rm submm})$, 
where $f_{ \rm AGN}$ is the luminosity-weighted AGN fraction.
\item The star formation rates in UV-bright galaxies could be 
underestimated because either the UV-luminosity-to-star-formation-rate 
conversion factor assumed is too large or the UV luminosities are 
underestimated due to internal absorption in the galaxies, as even small 
amounts of dust can have appreciable signatures at UV wavelengths. It is 
conventional to make corrections for this latter effect using an empirical
extinction law (Calzetti et al.~1997). However, recent near-infrared 
observations (Pettini et al.~2001) suggest that the star formation rate 
derived from the H$\beta$ line can be close to that derived from the UV 
luminosity with {\bf no} correction for extinction. 
\end{enumerate}

\section{$\gamma$-ray bursts and massive stars}

Until recently, GRBs were devoid of observable traces at other wavelengths. 
Measurement and localisation of fading X-ray signals from some GRBs 
has now lead to the detection of optical and radio afterglows, and enabled 
the measurement of redshifts and the identification of host galaxies. The 
bursts with known redshifts (in the range 0.43 $\le \,z \, \le$ 4.5) are brightenough to be detectable out to much larger distances than those of the most
luminous quasars. An example of this was the discovery (Akerlof 
et al.~1999) of a prompt, extremely bright ($V=9$), optical flash 50\,s after 
GRB 990123 started at $z=1.6$.  However, such bright prompt optical
flashes may be very rare, because they have not yet been detected in
another burst. Within the first minutes to hours after the burst,
optical light from the afterglows of bursts normally has $10 < V < 15$.\\

One can understand the dynamics of the afterglows of GRB in a fairly
simple manner, independent of any uncertainties about the progenitor
systems, from the relativistic generalisation of the method used to
describe supernova remnants. The basic model for GRB afterglow
hydrodynamics is a relativistic blast wave expanding into the
surrounding medium \cite{Rees94,sari95,spn96,mes97}. 
External shocks arise from the
interaction of the shell with  the surrounding matter. The 
typical scale is the Sedov length, $l \equiv (E/n_{\rm ism}
m_{\rm p}c^2)^{1/3}$. The rest-mass energy $E$ within a sphere of
radius $l$ equals the energy of 
the shell. Typically $l\sim 10^{18}$cm. The  relativistic external shocks 
convert a significant fraction of their kinetic energy into radiation at
\begin{equation}
R_{\rm aft}=l/\gamma^{2/3} \approx 10^{-2}-10^{-1} {\rm pc}, 
\end{equation}
the radius at which the external mass of ISM encountered equals
$\gamma^{-1}$ of the mass of the shell.\\  

The most widely discussed models of the central engine of GRBs involve
either the cataclysmic 
collapse of a massive star in a very energetic hypernova-like explosion
\cite{pac98,mw99,mc01}, or the coalescence of two compact objects such as
black holes or neutron stars \cite{lat76}\footnote{The formation of a
black hole with a debris torus around it is a crucial ingredient of
both these scenarios. The binding energy of the orbiting debris and the spin
energy of the black hole are the two main energy reservoirs available, with 
the total extractable energy being up to $10^{54}$ ergs \cite{Rees99}.}. The 
former are expected to occur near to or within star-forming regions of 
their host galaxies, while most of the latter are expected to occur outside the
galaxies in which they were born.\\ 

For the long-duration GRB afterglows localised so far, the host galaxies 
show signs of the ongoing star-formation activity necessary for the 
presence of young, massive progenitor stars (e.g. Paczy\'{n}ski 1998; 
Fruchter et al. 1999). Castander \& Lamb (1998) showed that the light from 
the host galaxy of GRB 970228, the first burst for which an afterglow was
detected, is very blue.  Subsequent analyses of the colour of this 
(Castander \& Lamb 1999) and other host galaxies (Kulkarni et al. 1998; 
Fruchter 1999) have strengthened this conclusion, as has the detection of 
O[II] and Ly${\alpha}$ emission lines from several other host galaxies (see 
Bloom et al.~1998; Kulkarni et al.~1998). X-ray spectral edges and 
resonance absorption lines are expected to  be detectable in the immediate 
environment of the GRB, in particular from the remnants of massive 
progenitor stellar systems (M\'esz\'aros \& Rees 1998). Observations with 
the {\it BeppoSAX} and {\it Chandra} X-ray spectrographic 
cameras have provided tentative evidence for iron line and edge
features in the spectrum of at least two bursts (GRB 991216, Piro et
al. 2000; GRB990705, Amati et al. 2000). One possible explanation of
the iron lines is that X-rays from the afterglow illuminated an
iron-enriched supernova remnant, produced days or weeks before the
burst by the same progenitor, leading
to  iron  recombination line emission (Vietri et al. 2000;  but see
Rees \& M\'esz\'aros 2000 for an 
alternative explanation involving continued decaying X-ray emission
from the GRB outflow impacting on the stellar envelope following core
collapse). 
There is independent evidence
that suggests that at least in some bursts, a supernova may be  the
origin of  a red component, a factor of about 60 times brighter in
flux than the  extrapolated afterglow, detected several weeks after the burst
(GRB980326, Bloom et al. 1999; GRB970228, Reichart 1999).  The
presence of iron line features would strongly suggest a massive stellar
progenitor; however, the exact nature of the central engine and the 
progenitor depend on the details of the models, since we are not able
to observe them directly.\\   

These associations
provide indirect evidence that long-duration GRBs and their afterglows
are produced by highly relativistic jets emitted in core-collapse
explosions. In this context,
high-redshift GRBs could become invaluable probes of the history of
high-mass star formation and galaxy evolution (Blain \&
Natarajan 2000; Lamb \& Reichart 2000). 
Because $\gamma$-rays are not attenuated by
intervening dust and gas, the selection of cosmic sites of massive star
formation by this method is unlikely to be affected by the biases associated
with dust absorption. 

\section{Extinction of GRB afterglows}

Because GRBs represent the final phase of the evolution of very massive
stars which do not live long enough to leave their birth place, 
a large fraction are likely to be still enshrouded in their
placental clouds of molecular gas and dust. This environment  
will attenuate the GRB afterglow
radiation at X-ray and optical wavelengths.  Since we expect the
molecular clouds and dust to be more prevalent in submm galaxies than
in UV ones, we expect the afterglows originating in the submm
galaxies to be more attenuated.

\subsection{Optical extinction}

\begin{figure}
\begin{center}
\vskip-4mm
\psfig{file=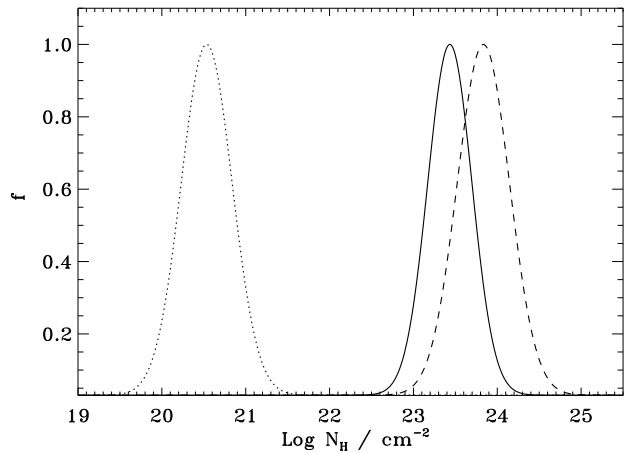, width=8.65cm}
\end{center}
\vskip-2mm
\caption{
Distributions (normalised to unity at the peak) of column densities
of lines of sight towards stars forming in an idealised compact
spherical gas cloud
with the mass and size of the infrared source in Arp 220 (see Fig.~1b;
Solomon et al.~1997).  The solid and dashed lines represent distributions for
two different density distributions of stars, gas, and dust
$\rho (r) \propto r^n$ within the cloud, with $n=0$ and -2 respectively.  
The much lower Galactic column density distribution for the sample of 
observed bursts without optical afterglows (Lazzati et al.~2001) are
represented by the dotted line.}
\end{figure}

Dust that is associated with a column density $N_H$ of cold gas 
extinguishes $V$-band optical radiation by an amount
\begin{equation}
A_V = N_{\rm H} / 1.79 \times 10^{21} {\rm cm}^{-2},
\end{equation}
if external galaxies have dust-to-gas
ratios similar to that of the Milky Way (Predehl \& Schmitt 1995).

The line-of-sight extinction to a GRB is, however, complicated by local 
physical processes.  For example, optical/UV/X-ray radiation from the GRB 
may heat and sublime nearby dust (Waxman \& Draine 2000), or winds
from the GRB progenitor may tunnel a hole through the  
surrounding medium. A slowly-varying near- and
mid-infrared emission should be detectable from the dust reprocessed 
in the vicinity of the GRB (Venemans \& Blain
2001), especially if the  
GRB emission is isotropic. Both processes would partially allow the escape of 
optical photons which would otherwise be extinguished. Such processes 
are probably important as (i) the observed optical absorption is 10--100 
times less than that inferred from the corresponding X-ray absorption, and 
(ii) the standard afterglow model fits to the data seems to indicate an 
ambient gas density that is generally lower than that expected in
star-forming clouds (Galama \& Wijers 2001). The result is that optical 
afterglows could be visible at distances far from the GRB in the absence of 
further extinction.\\ 

In ULIGs, however, there {\bf is} expected to be very substantial
further extinction along most lines of sight.  There can be up to tens
of magnitudes of extinction, maybe more, to the centre of ULIGs at
optical wavelengths, as shown in Figure 4 for the case of Arp 220, where
considerable amounts of dust are certainly present.

This cumulative extinction is due to the large
amounts of dust along our line of sight and differs fundamentally from
the extinction removed  by the sublimation of dust, which is
essentially a local phenomenon.
The X-rays from the GRB are unable to destroy the
dust responsible for the bulk of the extinction in ULIGs, since the
X-ray/UV flux $\sim d^{-2}$, so dust grains at $>>$ 10 pc from the GRB are
unaffected (Waxman \& Draine 2000).

\begin{figure}
\begin{center}
\vskip-4mm
\psfig{file=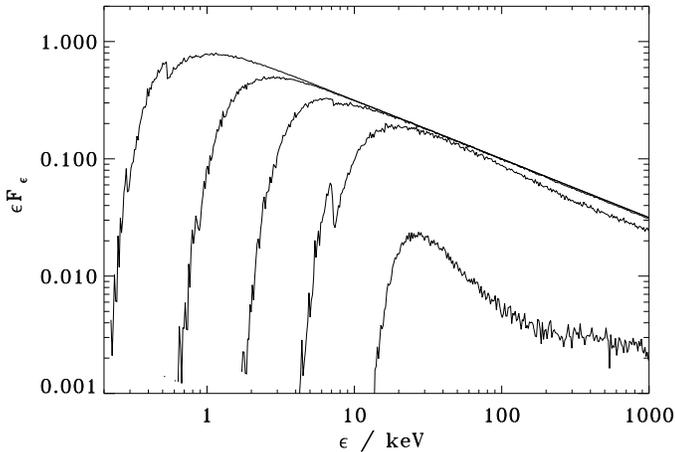, width=8.65cm}
\end{center}
\vskip-4mm
\caption{
The flux that would be received by the observer (in arbitrary units) for a
power-law background source with photon index 2.5
(GRB 980519; Nicastro et al.~1999)
through increasing column density (from left to right) 
$N_{\rm H}=$ 10$^{21}$, 10$^{22}$, 10$^{23}$, 10$^{24}$ and 10$^{25}$\, 
cm$^{-2}$. The effects of both Compton downscattering and photoelectric 
absorption have been included. Solar metallicity is assumed.}
\end{figure}

\subsection{High-energy extinction}

The X-ray radiation transmitted through high atomic column densities is 
modified significantly by both absorption and Compton scattering. To study 
the effects we consider a homogeneous spherical cloud of cold material
surrounding a central source of X-ray photons. We use the Monte Carlo code 
constructed by Wilman \& Fabian (1999), which computes the transmission 
through a sphere with a radius set by the column density $N_{\rm H}$. Both 
photoelectric absorption by neutral material (of solar composition) and 
Compton down-scattering are modelled, the latter being described by the
Klein--Nishina cross section, with elastic
scattering enhancing the photo-electric absorption probability (see
Madau, Ghisellini \& Fabian 1993). The downscattering of X-ray
photons by cold electrons modifies the transmitted spectrum above
an energy $E \sim m_{\rm e}c^2/\tau_{\rm C}^2$.
The results of some Monte Carlo
calculations are shown in Fig.~5, where an incident power law spectrum with
photon index 2.5 is seen in transmission through clouds of different column
density $N_H$. Only 40--70 per cent of the power
incident in the range 50--400 keV directly escapes from a medium
with $\tau_{\rm C}=$3--5; the range of optical depths $\tau_{\rm C}=1$--6
corresponds to $N_{\rm H}=10^{24.2 - 25}$ cm$^{-2}$. This percentage
progressively increases in the $\gamma$-ray region due to the
Klein--Nishina reduction of the Compton cross section at high energies.\\

\section{Observational constraints}

\subsection{Optical emission}

About three quarters of the star formation
in the Universe appears to be heavily obscured (see Section 2). If the
SCUBA galaxies are  
high-redshift counterparts of the ULIGs, and are powered mainly by
star-formation, then one would expect 
three quarters of GRBs to have no optical afterglows. 

The optical observations of GRB
afterglows are marginally consistent with this picture.
Optical afterglows have been detected for about one third of the
well and rapidly localized GRBs. Most of
these bursts, however,  were observed
with different telescopes, at various times and in different filters.
Among the 31 X-ray
afterglows detected by the WFC of {\it BeppoSAX}, excluding GRB 980425 and
X-ray transients, ${60 \pm 15}$\% had no optical
counterparts. Lazzati, Covino \& Ghisellini (2001) demonstrate that this is
due neither to adverse observing conditions nor delays in performing the
observations. Some limitations are necessarily inherent in their 
approach and selection of data. Nevertheless, recent studies using
different  burst samples give similar results: Reichart \& Yost 
(2001) find that   58 $\pm 13$ of all bursts (95 \% of all ``failed''
optical afterglows preceding GRB 000630)  have no
optical counterpart down to R=24; while  Djorgovski et al. (2001)
show that the maximum fraction of optical afterglows hidden by dust
is of the order of 50  $\pm 10$. 

Lazzati et al.~(2001) also find that X-ray transients
with no optical counterparts are not affected by unusually large
Galactic absorbing columns. They conclude that a minimum
absorption of 2 magnitudes in the R band is required for
more than half of the bursts\footnote{To convert dust absorption
values from a wavelength to a different one, Lazzati et al.~(2001) use
the analytic approximation for the dust extinction curve given in
Cardelli et al (1989), assuming that the bursts are at $z$=1, close
to the average value of the detected afterglows.}. This is much less
that the line-of-sight extinction expected in ULIGs (Fig. 4), but more
than would be expected for a GRB embedded in the Orion Nebula.\\

The non-detection of afterglows at optical/infrared wavelengths would
provide interesting insight into the location of obscured star
formation. If the number of bursts with no optical afterglows
is much less than three quarters, this could mean that these
GRBs did not occur in heavily optically thick molecular cores like those
observed in ULIGs such as  Arp 220 (see
Section 6 for further discussion).

\subsection{X-ray and $\gamma$-ray emission}

\begin{figure}
\begin{center}
\vskip-4mm
\psfig{file=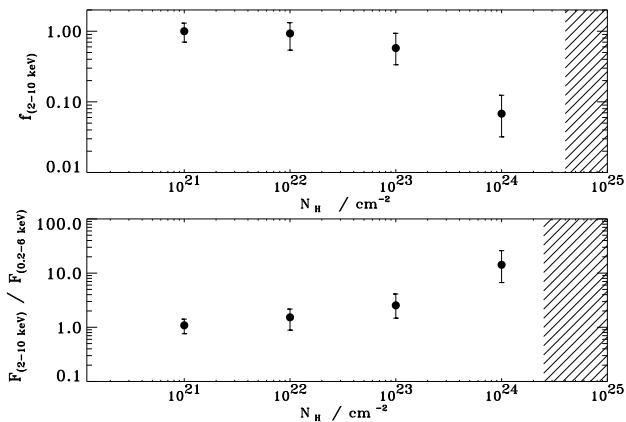, width=8.65cm}
\end{center}
\vskip-1mm
\caption{
Effects of absorption at X-ray wavelengths.
The upper panel shows the fraction of intrinsic flux 
expected in the {\it BeppoSAX} High Energy Channel (2--10\,keV) for
a source with a power-law spectrum of photon index 2.5
seen through obscuring material
of column density $N_{\rm H}$ . 
The lower panel shows the expected ratio of the emergent flux in the
{\it BeppoSAX}
Low Energy Channel (0.2--6\,keV) to the flux in the High Energy Channel
(2-10\,keV) as a function of $N_{\rm H}$.    
Solar metallicity is assumed.  The shaded area corresponds to complete
attenuation.}
\end{figure}

X-ray afterglows were observed in almost all the bursts detected by
the WFC of {\it BeppoSAX} (Lazzati et al.~2001). Even for extreme submm
galaxies like Arp 220, the obscuring column densities are not high
enough (Fig.~4) to  completely extinguish
the hard X-ray fluxes, and so $F_{\rm submm}$ equal to three quarters
is consistent with these observations as well.      

These results imply that even if GRBs are
located in the most absorbed regions of SCUBA galaxies, then the
corresponding absorption is not enough to hide their X-ray
afterglows. At these sorts of column densities, however, we expect the 
ratio between the {\it BeppoSAX} LECS (0.2--6\,keV)
and MECS (2--10\,keV) for bursts with
non-detected optical afterglows to be affected (Fig.\,6), with the
soft X-rays being considerably more attenuated than the hard ones.  

Column densities of the order of $10^{23.5}$\,cm$^{-2}$ or higher are 
required for noticeable effects on the X-ray spectrum; this corresponds to 
the average line-of-sight extinction to the core of Arp 220.  
The transmitted spectra in the 0.5--2\,keV energy range would be
almost completely extinguished; measurements in this energy interval are
therefore particularly important to probe the global column
density of the parent galaxy.
 
Interestingly, the X-ray
and $\gamma$-ray properties of  X-ray
transients with no optical counterpart are not systematically
different from those with an optical detection,
indicating that they are indeed intrinsically optically faint (Fynbo
et al. 2001; Lazzati et al.~2001). Based on the BATSE
Hardness Ratio (hereafter HR, defined as the ratio of the fluence at
100-300 keV and the fluence at 50-100 keV)
derived by Fynbo et al. (2001) for the GRB with
non-detected afterglows, we estimate that the line-of-sight column
densities for these bursts are characteristically below
$10^{23.5}$\,cm$^{-2}$, but greater than $10^{22}$\,cm$^{-2}$. 
These results assume that the sources are
located at $z$=1.0 and have a power-law spectrum. The column densities
derived here are much greater than  those typically  found in
star-forming clouds, but generally less than would be
expected for a GRB deeply embedded in the core of Arp 220. This
analysis agrees with the idea that the abscence of optical detections
is most likely the result of circumburst extinction. The density of the
burst medium probably spans many orders of magnitude from
densities as low as the ones found in typical star-forming clouds to
densities as high as the ones found in the core of Arp 220.

\subsection{Very high energy emission}

If the column density of the parent galaxy is greater than
$10^{25}$\,cm$^{-2}$ in any significant fraction of the SCUBA sources,
then we expect the GRB spectra  to peak at very high energies (Fig 6).  

GRBs are typically identified by sharp increases in the count rates in the 
50--200\,keV band. Although some of these bursts have been observed 
up to very high energies  (e.g., Atkins et al. 2000), concerns have been 
voiced that a whole class of very hard bursts may have been missed due to 
the use of the X-ray band for identifying GRBs (e.g. Piran \& Narayan 1996). 
If the current detection strategy discriminates against bursts with harder 
than average spectra, then there may exist GRBs that emit strongly at 
energies greater than 1\,MeV, but whose emission near 100\,keV falls 
below the detection threshold. The data show very small numbers of
hard bursts, e.g., only five bursts out of the 136 in the Cohen, Piran
\& Narayan (1998) sample are harder than 1 MeV. Nevertheless, this
does not mean that there are fewer GRBs above 1 MeV, since harder
bursts have fewer photons and thus the decrease of BATSE sensitivity above
300keV makes their detection very difficult (Cohen et al. 1998). The
existence of a large population of unobserved hard gamma-ray bursts
is still an open question. Harris \& Share (1998) used data 
from the Solar Maximum Mission Gamma-Ray Spectrometer to show that
there is a deficiency of GRBs with hardness above  3 MeV, relative to
those peaking at $\sim$ 0.5 MeV, but these data are consistent with a
population of hardness that extends up to 2 MeV.  If SCUBA galaxies
with column densities greater than  $10^{25}$\,cm$^{-2}$ exist in
large numbers, then a large population of unobserved hard gamma-ray
bursts may exist. These bursts can be used to detect a putative
population of highly obscured galaxies. 

\section{Caveats}

A number of caveats apply to our analysis.

\subsection{Massive stars  and  GRB formation}  
In the previous sections we have assumed that the underlying physics
which turns a massive star into a GRB is the same for
stars forming in both the UV and submm modes. 
This may be incorrect for various reasons.\\
 
Quantitative insight
into the formation of GRBs is hindered
by the lack of detailed core-collapse calculations.  However, the
insightful work of MacFadyen \& Woosley (1999) suggests that
GRBs can occur in stars with a range of radii. The radii depend on the
evolutionary state of the massive progenitor, its binarity, rotation, and more
importantly metallicity \cite{mw99,heger00}. A picture in which 
stars produced in the submm mode are more metal rich than those
in the UV mode would lead to
an overestimate of $F_{\rm submm}$ via GRBs
since low metallicity raises the
mass threshold for the removal of the hydrogen envelope by stellar winds,
thus increasing the mass of the heaviest helium core and favouring black
hole formation \cite{m91,mw99,mm00}. 
In the local Universe, UV-bright star-forming galaxies tend to have
very low metallicities (e.g.~I Zw 18; Hunter \& Thronson 1995)
relative to submm-bright ULIGs (Sanders \& Mirabel 1996). The same may
be true at high redshift, with stars in submm-bright galaxies probably 
forming at higher densities than the stars in UV-bright galaxies, which may 
also affect the evolutionary state of the progenitors to GRBs. One expects 
various outcomes ranging from GRBs with large energies and durations, to 
asymmetric, energetic supernovae with weak GRBs. Observations of the 
medium surrounding a GRB would distinguish naturally between different 
stellar explosions \cite{che99,ram01}.

\subsection{Extinction in the UV mode}
In the previous sections we have assumed that the sources of UV 
star-formation are not highly obscured and therefore that the afterglows 
of GRBs in this mode are not likely to be attenuated. However, there
may be patches of dust which could attenuate optical but not
X-ray afterglows, if their column densities are between 10$^{21}$ and 
10$^{23}$\,cm$^{-2}$. Such patches would need to have greater column 
densities than lines of sight through the Orion Nebula (Lazzati et al.~2001), 
but less than those through Arp 220. This implies that GRBs in UV-bright
galaxies could also be attenuated by dust, biasing our estimate of 
$F_{\rm submm}$ too high. 
 
If {\bf all} GRBs ultimately turn out to have optical afterglows, and
$F_{\rm submm}$ is intrinsically low, then this concern is not longer 
valid.  We 
will then also learn that there is not a large  amount of local
extinction in UV-bright galaxies. Such a possibility is suggested by
the approximate concordance  between the shape of the UV Madau plot at
$z<3$ and the GRB rate derived  using the luminosity--variability
relation (Ramirez-Ruiz et al.~2001).  

\subsection{AGN fraction}
It is worth examining the assertion that the submm-bright sources are 
powered by star formation rather than dust-enshrouded AGNs.
The arguments in favour of star formation as the dominant power source 
are
\begin{enumerate} 
\item  
The radio luminosities of the SCUBA sources (Smail et al.~2000) are not 
large and are consistent with having been produced by supernova remnants.
\item 
Radiative transfer modelling (Rowan-Robinson 2000 \& references
therein) suggests that most of the far-infrared radiation emerging longward 
of 50\,$\mu$m in the rest frame, at which the bulk of the energy from the
SCUBA sources emerges, originates from star formation as opposed to an 
AGN. However, the uncertain and probably complex geometry of the 
SCUBA galaxies could complicate this interpretation.  
\item 
Most luminous SCUBA sources are not luminous X-ray sources 
and so are consistent with being starburst-powered. This deduction is,
however, not conclusive. Hard X-ray observations can uncover the
presence of a dominant AGN power source in even highly obscured sources,
and hence provide a
simple estimate of the AGN fraction in the submm mode (Bautz et al.\
2000; Fabian et al.\ 2000). Unfortunately, even for high-redshift sources 
{\it Chandra} provides relatively poor hard X-ray response, and so these 
X-ray studies have so far only confirmed the presence of 
modestly-obscured AGN in two galaxies already confirmed to contain AGN 
from optical spectroscopy.
\end{enumerate}

The primary argument against star formation as the dominant power 
source is that any stars produced in the SCUBA galaxies need to disappear 
by the present epoch, as only a negligible fraction (about 1 per cent; 
Trentham 2001) of the stars in the local Universe are found at the high 
densities observed in ULIGs and thought to exist in the SCUBA galaxies 
($>$100\,M$_{\rm \odot}$\,pc$^{-3}$). AGNs then become an attractive 
alternative since super-massive black holes exist in just the right number 
to have been produced in the SCUBA galaxies, given the energy output from 
these galaxies: see both Table 2 and the accompanying discussion in 
Trentham \& Blain (2001).

\subsection{Stellar IMF}
The precise form of the IMF in starburst galaxies is very much an open
question. Some indirect observations that guide us to believe that the IMF 
in the local ULIGs could be high-mass biased (e.g.~Goldader et al.~1997).
If the analogy between these objects and high-redshift submm galaxies
is valid, then the IMF could also be biased in the high-redshift galaxies, 
leading to an overestimate of $F_{\rm submm}$ in Section 2. 

Another important issue is the relative shapes of the very high-mass-end
of the IMF in submm and UV-bright galaxies, since very high
mass stars are thought to produce GRBs.  
Any difference of this type could strongly affect a comparison of
$F_{\rm submm}$ with the fraction of GRBs without afterglows. 
A strong difference of this type, however, is unlikely in the nearby
Universe,  since the radio--far-infrared correlation (Condon, Frayer \& 
Broderick 1991) holds over a large range of galaxy luminosity and the 
very-high mass stars are the ones responsible for generating the 
far-infrared radiation.  First indications (Frayer et al.~1998, 1999; Ivison et
al. 1998, 2000) are that the few high-redshift SCUBA sources that have been 
studied in detail lie close to the low-redshift radio--far-infrared
correlation. 
 
\subsection{Alternative causes of non-detection of optical afterglows}
We have assumed that dust obscuration is the most likely reason for
the non-detection of optical transients. However, there are other
possible explanations:

\begin{enumerate}
\item 
Bursts with and without afterglows are
intrinsically different -- bursts of a given
$\gamma$-ray luminosity could be either optically bright or faint 
(Fynbo et al. 2001; Lazzati et al.~2001)
\item 
The absence of detectable
optical flux accompanying strong X-ray emission in the 2--10\,keV {\it
BeppoSAX}  energy band may simply be due to the strong redshift dependence
of dust obscuration at optical wavelengths in comparison to the
X-rays, i.e. $\tau_{{\rm opt}} / \tau_{{\rm x-ray}} \propto (1 + z)^4$
(Taylor et al. 1998).
\item 
A large fraction of GRB occur at $z$ $>>$ 5 and so are optically invisible
due to Lyman-$\alpha$ absorption by intervening
systems (Fynbo et al. 2001; Lazzati et al.~2001). 
However, using the sample of {\it BATSE} bursts without optical
afterglows for which high-resolution light curves are 
available, we can estimate the redshifts using the variability--luminosity
relation derived by Fenimore \& Ramirez-Ruiz
(2001) for long-duration GRBs; all are less than 5.
\end{enumerate}
 
\section{Summary and future prospects }

In summary, we expect about three quarters of the star formation in the
Universe to have occurred in submm-bright as opposed to
UV-bright galaxies.  High-redshift submm galaxies (the SCUBA galaxies)
are thought to have properties similar to local ULIGs, based on the
concordance between their luminosities and spectral energy distributions.
If this is the case, we expect that GRBs in submm-bright
galaxies should typically have their optical afterglows absorbed by internal
dust extinction, but only very few should have had their 2$-$10 keV
X-ray afterglows completely extinguished.  

The value of three quarters that we quote is marginally consistent with 
observations of GRBs: $60 \pm 15$ per cent of GRBs have no detected 
optical afterglow, although  almost all have an X-ray afterglow.

\begin{figure}
\begin{center}
\vskip-4mm
\psfig{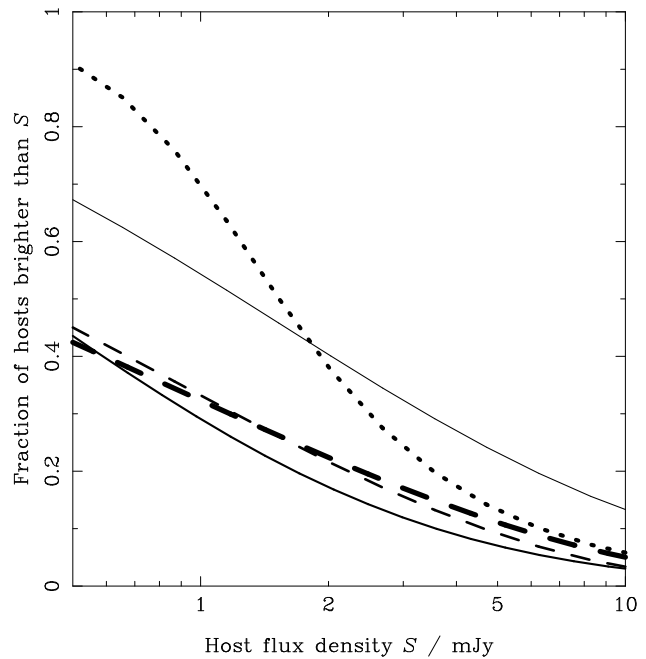}
\end{center}
\vskip-1mm
\caption{The predicted fraction of submm-luminous host galaxies of GRBs 
that exceed a certain threshold in 850-$\mu$m flux density
based on five different models of the star-formation history, that 
fit the results of far-infrared and submm observations. The thin 
solid and dashed lines show the results of Blain et al.\ (1999b,c)
respectively. The thick solid and dashed lines show the equivalent 
updated results described in Blain (2001). The dotted line shows the result 
of Barger, Cowie \& Sanders 1999. The Blain (2001) results are likely
to be most  reliable, and indicate that of order of 20 per cent of GRB
host galaxies might  be expected to exceed the 2-mJy 850-$\mu$m
detection threshold of the  
SCUBA camera at the JCMT, which can be achieved in several hours of 
integration in photometry mode. The future ALMA interferometer array 
will be able to make such detections, and to produce 
sub-arcsecond-resolution images of the hosts in a few seconds of 
integration.}  
\end{figure}

A more definitive statement could be made with detailed  
observations of soft X-ray afterglows
(e.g. {\it HETE-2} and {\it Swift}). This is because we
expect X-ray afterglows to be extinguished in the 0.5$-$2 keV, but not at
harder X-ray bands (Fig. 5). If X-ray afterglows of a large number of GRBs
disappear, then this would provide strong evidence that much of the
star-formation activity in the Universe happened in
a heavily obscured  submm mode. 

The detection of afterglows at  
soft X-ray wavelengths would then provide interesting insight into the
location of cosmic star formation. This could mean that our value of
three quarters is too high, 
perhaps because $f_{\rm AGN}$ is significant, or one of the other caveats
discussed in Section 6 is important. Alternatively, it might mean that
star formation in submm-bright galaxies does 
not occur in very compact regions, and so the analogy with local
ULIGs is incorrect -- the luminosity and spectral energy
distribution concordance would then be just a coincidence. 
This would also resolve a problem discussed in Section 6.3 -- if 
stars form in submm-bright galaxies away from
very compact regions, then most of the stars in the local
Universe could form in this way, and not just the 1 per cent with
high densities in elliptical galaxy cores.  
Looking for high-density molecular tracers like CS and HCN in 
high-redshift submm-bright galaxies would be one way to 
test the connection hypothesis (Trentham 2001). However,
such measurements would be difficult. \\

If GRB afterglow and submm
source positions can be correlated, then this would provide
a direct way to probe the nature of high-mass star formation
(Mirabel, Sanders \& Le
Floch 2001). At present, we  expect about 20 per cent of GRB host
galaxies to exceed the 2-mJy 850-$\mu$m detection
threshold of the SCUBA camera at the JCMT if the GRB rate  follows the
cosmic star formation rate (Fig. 7).  An important caveat is
that most detected X-ray afterglows had error boxes of a few square
arcminutes, a region large enough to contain from tens up to a few
hundred galaxies. Therefore it is very difficult to identify the host galaxy 
unambigously without detections of optical, infrared, radio or {\it 
Chandra} X-ray afterglows. 

Studies of the galaxy hosts of the GRBs detected with the next generation of 
$\gamma$-ray observatories (e.g. {\it HETE-2} and {\it Swift}) with
submm and infrared telescopes (e.g. {\it SIRTF}, {\it NGST}, ALMA and
{\it FIRST-Herschel}), would reveal whether or not most of the
high-mass star formation activity in the Universe is heavily obscured.

\section*{Acknowledgements}
We thank D. Lazzati, M. Rees, A. Merloni, E. Rossi, I. Smail, N. Tanvir and 
A. Emmanouil for useful
comments and suggestions. We are particularly grateful to  R. Wilman
for useful insight regarding the Monte-Carlo calculations. We thank
I. Smail, R. Ivison and M . Pettini  
for providing the images. We thank J. Greiner for maintaining the GRB 
web archive. ERR acknowledges support from CONACYT, SEP and the ORS 
foundation. AWB thanks the Raymond \& Beverly Sackler Foundation for 
support.

\end{document}